\documentclass[aps,pra,superscriptaddress,twocolumn,10pt]{revtex4-1}

\usepackage{ucs}
\usepackage[latin1]{inputenc}
\usepackage[english]{babel}
\usepackage[usenames, dvipsnames]{xcolor}
\usepackage{amsmath, 
			amsfonts,
            booktabs,
            microtype,
            mathtools, 
            graphicx,
            color,
            nicefrac,
            url, 
            placeins,	
			braket,
			appendix} 	

\begin{document}
\title{A versatile, high-power 460\,nm laser system for Rydberg excitation of ultracold potassium}
\author{Alda Arias}
\affiliation{Physikalisches Institut, Universit\"at Heidelberg, Im Neuenheimer Feld 226, 69120 Heidelberg, Germany.}
\author{Stephan Helmrich}
\affiliation{Physikalisches Institut, Universit\"at Heidelberg, Im Neuenheimer Feld 226, 69120 Heidelberg, Germany.}
\author{Christoph Schweiger}
\affiliation{Physikalisches Institut, Universit\"at Heidelberg, Im Neuenheimer Feld 226, 69120 Heidelberg, Germany.}
\author{Lynton Ardizzone}
\affiliation{Physikalisches Institut, Universit\"at Heidelberg, Im Neuenheimer Feld 226, 69120 Heidelberg, Germany.}
\author{Graham Lochead}
\affiliation{Physikalisches Institut, Universit\"at Heidelberg, Im Neuenheimer Feld 226, 69120 Heidelberg, Germany.}
\author{Shannon Whitlock}
\affiliation{Physikalisches Institut, Universit\"at Heidelberg, Im Neuenheimer Feld 226, 69120 Heidelberg, Germany.}
\affiliation{IPCMS (UMR 7504) and ISIS (UMR 7006), Universit\'e de Strasbourg and CNRS, 67000 Strasbourg, France.}
\date{\today}
	
\begin{abstract}
We present a versatile laser system which provides more than 1.5\,W of narrowband light, tunable in the range from 455--463\,nm. It consists of a commercial Titanium-Sapphire laser which is frequency doubled using resonant cavity second harmonic generation and stabilized to an external reference cavity. We demonstrate a wide wavelength tuning range combined with a narrow linewidth and low intensity noise. This laser system is ideally suited for atomic physics experiments such as two-photon excitation of Rydberg states of potassium atoms with principal quantum numbers $n>18$. To demonstrate this we perform two-photon spectroscopy on ultracold potassium gases in which we observe an electromagnetically induced transparency resonance corresponding to the $35s_{1/2}$ state and verify the long-term stability of the laser system. Additionally, by performing spectroscopy in a magneto-optical trap we observe strong loss features corresponding to the excitation of $s,p,d$ and higher-$l$ states accessible due to a small electric field.
\end{abstract}
	
\maketitle

\subsection{Introduction} 

The coupling of atomic systems with laser fields enables the extremely precise creation and study of model quantum systems, such as atom-light interactions~\cite{Cohen-Tannoudji1998}, cavity quantum electrodynamics~\cite{mabuchi2002cavity,haroche2013nobel}, optical atomic clocks~\cite{Ye2008,ludlow2015optical}, strongly correlated matter~\cite{bloch2008many,lewenstein2007ultracold}, and quantum simulators~\cite{Bloch2012}. Success in these areas is largely attributed to the development of techniques such as laser cooling and trapping which have enabled an amazing level of control over essentially all ground state properties of ultracold atoms, including their atomic motion, spatial geometry, coherence and interaction properties. At the same time, the continuing development of new laser sources capable of spanning almost the complete wavelength range (from ultraviolet to infrared), have made it possible to achieve quantum control over the single-atom electronic properties of many atoms, including their highly excited Rydberg states. Laser control of Rydberg states provides a novel platform for quantum science and technology, including: probing surface fields~\cite{tauschinsky2010spatially,abel2011electrometry,hattermann2012detrimental,thiele2015imaging}, sensing using thermal vapours~\cite{bason2010enhanced,Sedlacek2012,Fan2015}, quantum enhanced metrology~\cite{bouchoule2002spin,Gil2014}, quantum information processing~\cite{Saffman2010}, extreme nonlinear via electromagnetically induced transparency~\cite{pritchard2013nonlinear,Firstenberg2016} and driven-dissipative systems~\cite{lee2012collective,Glaetzle2012driven,ates2012dynamical,carr2013nonequilibrium, schempp2014full, Malossi2014, urvoy2015strongly,valado2016experimental,helmrich2016scaling}.

One outstanding application of Rydberg atoms is to exploit strong light-matter interactions to realize new types of quantum fluids enhanced by long-range interactions (Rydberg dressing)~\cite{Henkel2010,Pupillo2010,Cinti2010,Honer2010,Balewski2014,Gaul2015,Jau2015,Zeiher2016,Plodzien2016}. The challenge lies in reaching a regime of strong Rydberg atom-light coupling while at the same time minimizing decoherence.  Most cold atom experiments working with highly excited Rydberg states are realized using alkali atoms, in which the Rydberg state is accessed via a two-photon transition involving two laser fields (commonly in the infrared and blue wavelength ranges). In this case the relevant figure of merit for Rydberg dressing scales proportionally to the atom-light coupling to the Rydberg state and inversely with the effective decoherence rate~\cite{Helmrich2016}. Spontaneous emission from the short-lived intermediate state and dephasing due to laser frequency noise is primarily responsible for decoherence. Therefore, experiments require high intensities for the blue light (to compensate for the small transition dipole moments between low-lying and excited states) combined with narrow laser linewidths. A common approach involves the use of frequency-doubled semiconductor laser systems which can reach a wavelength tuning range of $8\,$nm, output powers $\lesssim 1\,$W and locked laser linewidths $\lesssim 200\,$kHz~\cite{Koglbauer2011,Pagett2016,eismann2016active}. 

\begin{figure}[t!]
	\centering
	\includegraphics[width = 1\columnwidth]{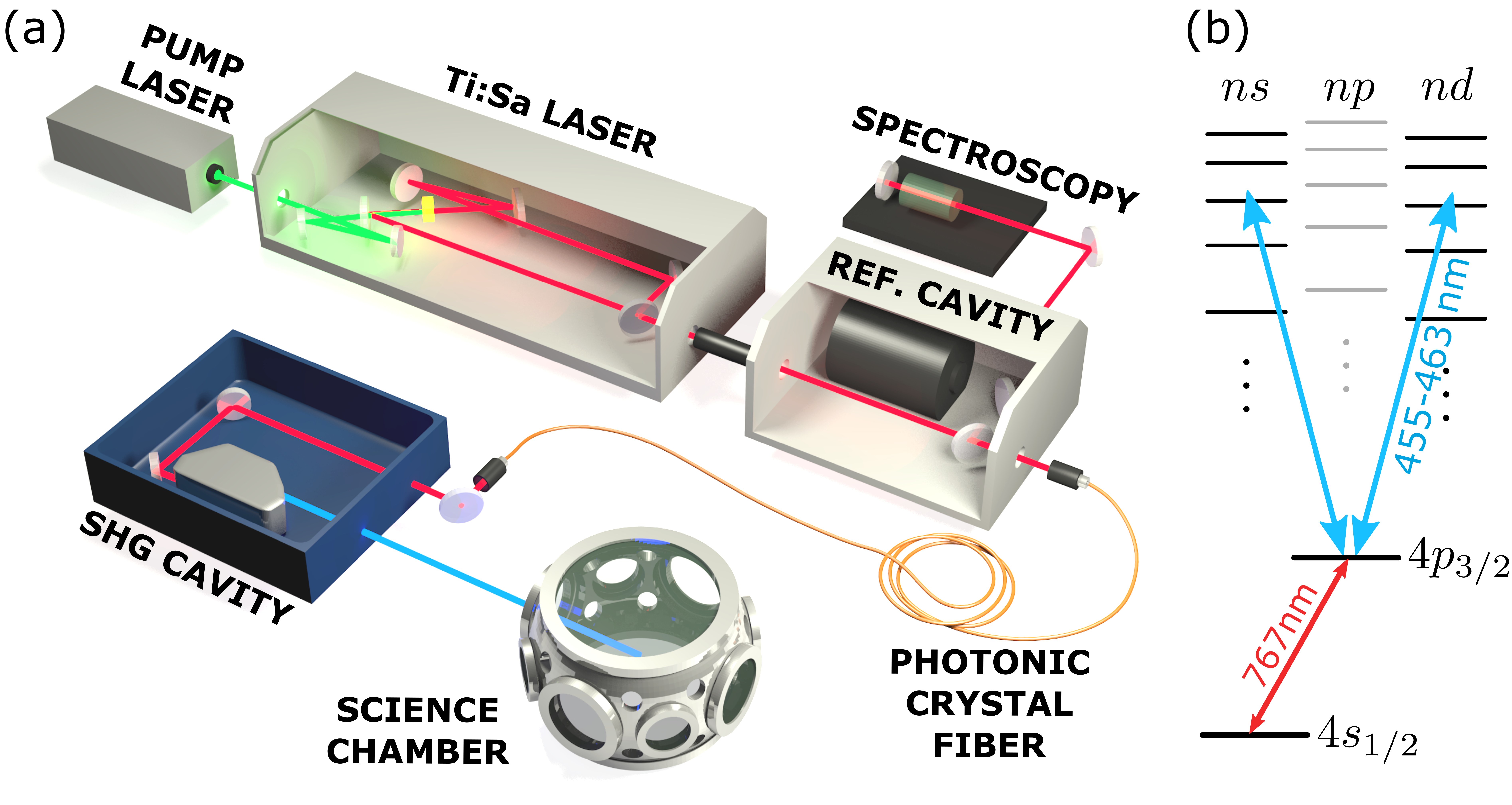}
	\caption{Schematic overview of the laser system and potassium level scheme used for two-photon Rydberg excitation. (a) The laser system consists of a Titanium-Sapphire laser and resonant-cavity frequency doubling (SHG). To improve the frequency stability on short and long timescales the Ti:Sa laser is locked to a reference cavity which in turn is stabilized to an absolute frequency via Doppler free spectroscopy of $^{39}$K atoms. (b) level scheme of the two-photon excitation of potassium to Rydberg $ns$ and $nd$ states. The laser system presented in here corresponds to the upper transition around 460\,nm.}
	\label{fig:The setup}
\end{figure}
In this paper we present an alternative high power, widely tunable and stable laser system (Fig.~\ref{fig:The setup}a) applied as the second step of a two-photon Rydberg excitation scheme of ultracold potassium.  In the case of potassium atoms, the typical relevant transitions are the $4s_{1/2}\rightarrow 4p_{3/2}$ and $4p_{3/2}\rightarrow ns_{1/2}\,\mathrm{or}\,nd_{3/2,5/2}$ transitions at 767\,nm and around 460\,nm respectively (Fig.~\ref{fig:The setup}b).  Coincidentally, the upper transition corresponds also to the wavelength range used for laser cooling of strontium at 461\,nm~\cite{Katori1999}, and therefore this laser has applications beyond Rydberg physics, e.g. optical frequency metrology~\cite{Ye2008,ludlow2015optical} and quantum sensors~\cite{Sorrentino2009}. It consists of a frequency doubled Titanium-Sapphire laser (Ti:Sa) stabilized to a reference cavity. The reference cavity is, in turn, locked to an atomic reference via Doppler free modulation transfer spectroscopy of a $^{39}$K thermal vapour at 767\,nm. We characterize the system by its high output power, exceeding 1.5\,W, combined with single-frequency operation, narrow linewidth and low intensity noise. The wavelength tuning range spans 455\,nm to 463\,nm, corresponding to Rydberg-states with principal quantum numbers from $n=18$ to above the ionization limit. We demonstrate its suitability for Rydberg excitation via two-photon spectroscopy of ultracold atoms released from an optical dipole trap, with which we study the long term stability of the laser system, and via laser driven ionization of Rydberg states in a magneto-optical trap (MOT), where we observe the excitation spectrum $ns,np,nd$ and higher-$l$ states over a wide range of principal quantum numbers. 

\subsection{Overview of the laser system}
\begin{figure}[t!]
	\centering
	\includegraphics[width = 1\columnwidth]{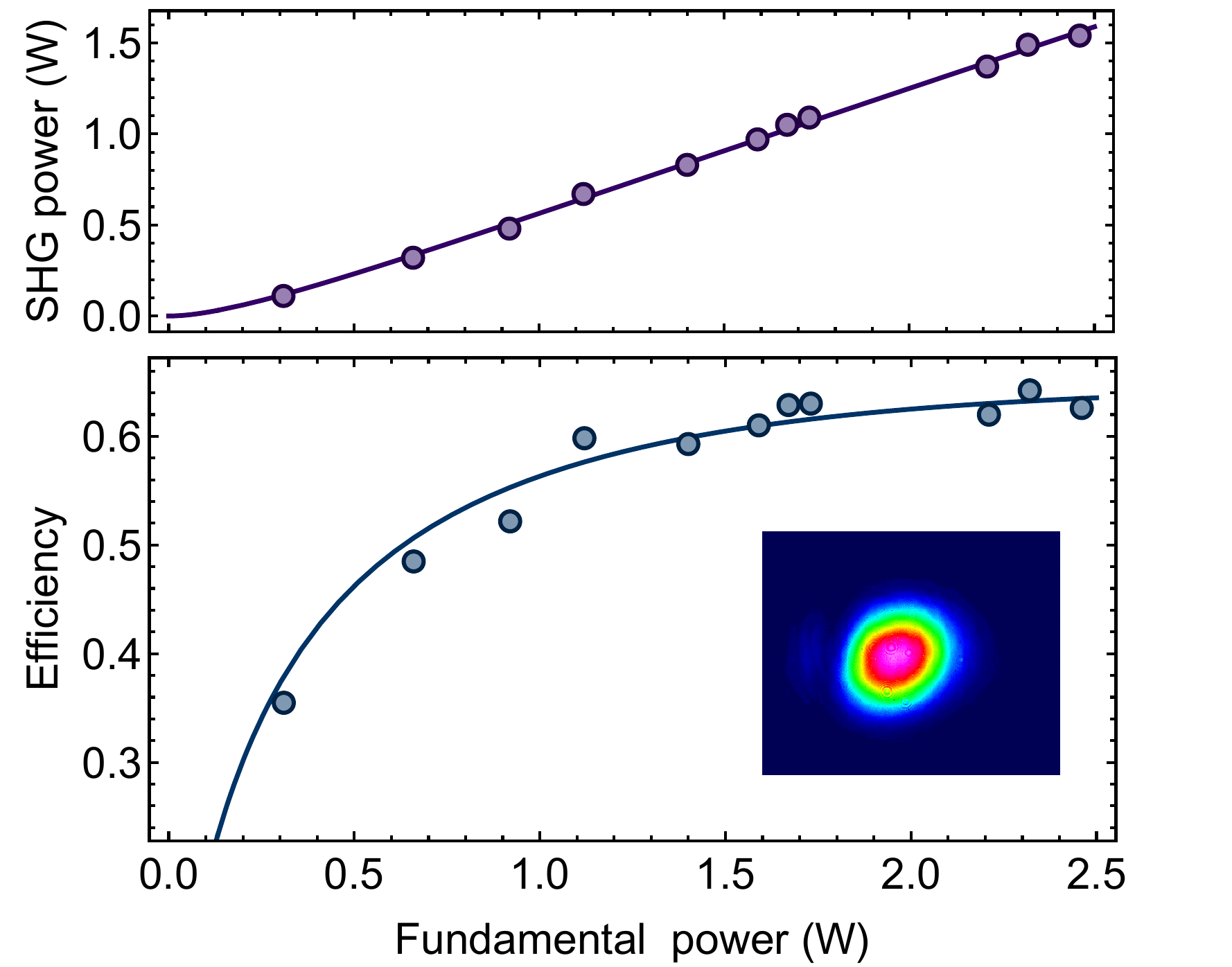}
	\caption{Performance of second harmonic generation (SHG). The top panel shows the SHG power as a function of the fundamental power at a wavelength of 456(912)\,nm, reaching a maximum power of 1.5\,W for an input of 2.5 W of the fundamental light. The bottom panel shows the SHG efficiency which saturates at around 60\% above 1\,W input power. The solid lines correspond to a model for the electromagnetic field propagation inside a Fabry-Perot cavity (details in the text). The inset shows a typical beam profile for the SHG cavity output beam.}
	\label{fig:SHGEfficiency}
\end{figure}

The complete laser system is composed of four main elements (Fig.~\ref{fig:The setup}a). A diode-pumped solid-state laser (Spectra-Physics Millenia eV) delivers up to 22\,W of light at a wavelength of 532\,nm. This laser serves as a pump for a Titanium-Sapphire (Ti:Sa) continuous wave ring-laser system (Sirah Lasertechnik Matisse TX) which we operate in the wavelength range between 910\,nm and 926\,nm. Frequency selection and single mode operation is achieved via a set of computer controlled motorized or piezo-driven frequency selective elements including a birefringent filter, a thick etalon and a thin etalon. A small fraction of the light is picked off to stabilize the laser frequency to an external reference cavity with a free-spectral-range of 1160\,MHz and a finesse of 250. The system optionally includes a nonresonant intra-cavity electro optical modulator (EOM) to allow high bandwidth feedback for laser frequency stabilization, however without it, it was possible to increase the output power by approximately 10\%, achieving typically 5\,W at 920~nm, without significantly influencing the measured linewidth.

The output of the Ti:Sa laser is transferred via an optical fiber to another laser table where the frequency doubling cavity and the ultracold atom apparatus is located. This enables additional flexibility for positioning the experiment and we found it to be more efficient than fiber coupling the frequency doubled light which is strongly affected by nonlinear scattering processes. To accommodate the high fundamental power we use a single mode polarization maintaining photonic crystal fiber (NKT-photonics LMA-PM-15) with which we achieve an overall fiber coupling efficiency of 50\%. After the fiber, the light is coupled into a commercial bow-tie ring cavity containing a nonlinear second-harmonic-generation (SHG) crystal (Toptica SHG-pro). More than 1.5\,W of single mode light around 460\,nm is then available to perform Rydberg excitation experiments.

\begin{figure}[t!]
	\centering
	\includegraphics[width =0.9\columnwidth]{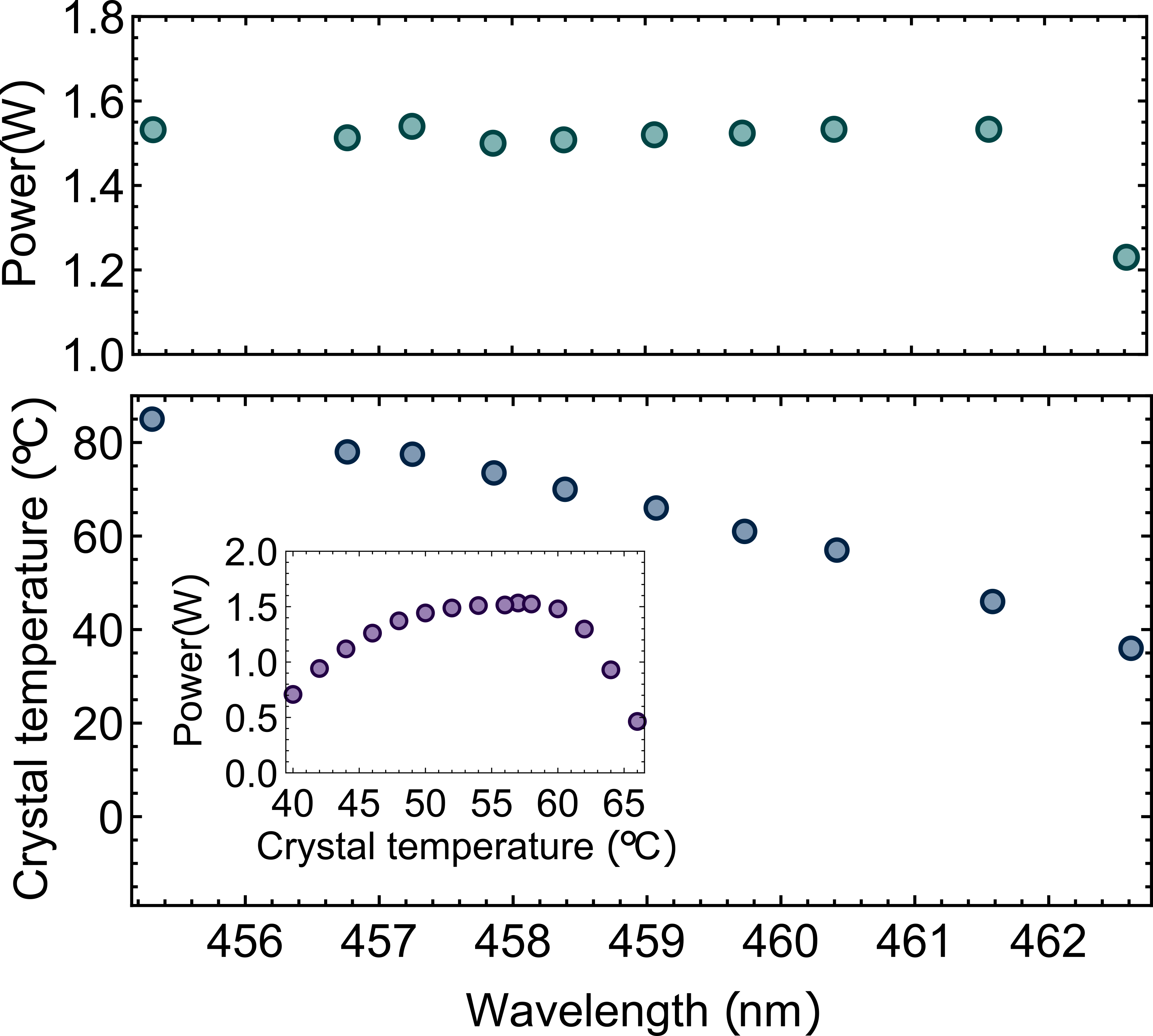}
	\caption{Wavelength tuning range of the frequency doubled laser spanning from 455 -- 463\,nm with a nominal output power $\approx$ 1.5W (upper panel). For each operating wavelength the optimal crystal temperature for phase matching must be found (lower panel). The inset shows the typical crystal temperature dependence of the output power.}
	\label{fig:CrystalTAndPower}
\end{figure}

\subsection{Characterization of the SHG output beam and wavelength tunability}
We have characterized the output of the SHG as a function of the fundamental power (Fig.~\ref{fig:SHGEfficiency}). The maximum output power has a quadratic dependence on the fundamental power below approximately 1\,W which crosses over to linear scaling at high powers. The crossover coincides with a saturation of the SHG conversion efficiency at around 60\% resulting in a maximum output power of 1.5~W blue light (for 2.5\,W fundamental). In comparable experiments without the photonic crystal fiber we achieved a maximum of 3\,W blue light (5\,W fundamental). This is substantially more than routinely operated amplified semiconductor diode-based systems \cite{Koglbauer2011,Pagett2016}. We found we can model the SHG performance by including the power absorbed to generate the second harmonic light as an additional loss term in the equations describing the electromagnetic field propagation inside a Fabry-Perot cavity (following Ref.~\cite{Hadjar00}) confirming that we reach close to the optimal operating conditions. To check the quality of the output beam we also measured the output beam profile at 1.5\,W output power which exhibits an approximately Gaussian profile with almost equal cross-sections in the transverse directions (figure \ref{fig:SHGEfficiency} inset).

Next we investigate the wavelength tuning range which was measured by varying the Ti:Sa wavelength and adjusting the SHG cavity and the crystal temperature to maximize the output power. Figure \ref{fig:CrystalTAndPower} shows the output power and optimal crystal temperature for phase matching measured over the wavelength range 455 - 463\,nm. A typical measurement of the output power as a function of temperature for a wavelength of 460.4\,nm is shown in the inset. The noticeable asymmetry can be explained as a consequence of crystal heating due to extra absorption away from the optimal phase matching condition. As the wavelength is reduced, higher crystal temperatures are required. We measured up to  $85\,^\circ$C which corresponds to a minimum wavelength of 455.2\,nm. Additionally, we confirm that an output power of $\approx 1.5\,$W can be achieved over almost the whole range of wavelengths (Fig.~\ref{fig:CrystalTAndPower}a). This setup has been used for more than one year with no sign of degradation and only requires minor adjustments from day-to-day when working with a fixed wavelength.

\begin{figure}[t!]
	\centering
	\includegraphics[width = 0.95\columnwidth]{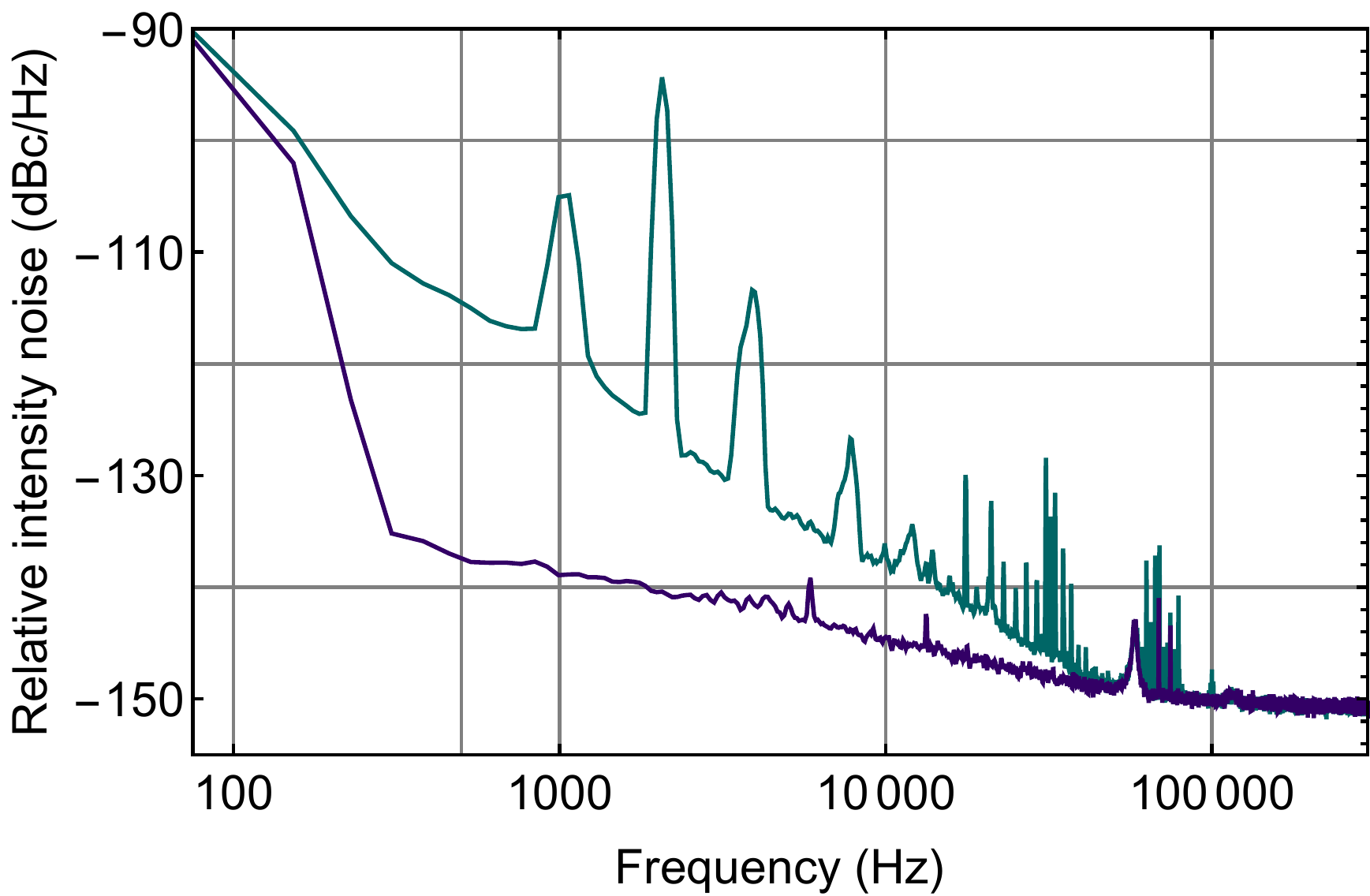}
	\caption{Relative intensity noise of the frequency doubled light over the bandwidth of 100 Hz to 5 MHz. The upper curve (green) corresponds to the laser intensity noise while the lower curve (purple) is the relative background noise level of the photodiode measured without incident light. The laser noise spectrum shows a 1/f behavior plus its characteristic noise spikes.}
	\label{fig:IntensityNoise}
\end{figure}

\subsection{Intensity noise characterization and laser frequency stabilization}

Short term fluctuations of the laser frequency and SHG cavity transmission are strongly suppressed by locking the Ti:Sa to an external reference cavity. To characterize this we measure the linewidth and intensity noise of the frequency doubled light. For linewidth measurements we use an optical spectrum analyzer (Sirah Lasertechnik, EagleEye) which can resolve frequency fluctuations as small as 20\,kHz using a lock-release-measure technique. The root-mean-square frequency noise is $115\pm 9\,$kHz, measured within a time window of $7\,\mu$s. The relatively short time window is chosen to reflect the typical excitation timescales in an ultracold atom experiment. The uncertainty is estimated from the root-mean-square deviation of several independent measurements taken over one hour. This is consistent with the Ti:Sa laser noise for which we measure a linewidth of $63\pm 5\,$kHz. Therefore, the whole laser system may be made spectrally narrower by locking the Ti:Sa to a reference cavity with a higher finesse.

In addition to laser frequency noise, it is important to minimize intensity noise which may cause spectral broadening or heating of the atoms due to AC Stark shifts during Rydberg excitation. Using a photo-diode and a spectrum analyzer we measure the relative intensity noise which is shown in Fig.~\ref{fig:IntensityNoise}(green line). The background noise level is measured without any light incident on the photodiode (Fig.~\ref{fig:IntensityNoise}, purple line). The spectrum exhibits an overall 1/f dependence with additional narrow resonances corresponding to different noise sources. For example, we determined that the two main peaks at 1\,kHz and 2\,kHz originate from the modulation of the fast piezoactuator used to lock the Etalon within the Ti:Sa cavity which could not be further reduced without impacting the frequency stability. Integrating the spectrum over the bandwidth range from 75\,Hz to 5\,MHz yields a root-mean-square noise level of less than 0.1 percent. 

While the reference cavity provides good short term stability we notice a typical drift of several MHz/hour under typical laboratory conditions. This is largely due to external temperature fluctuations as well as drifts associated with the piezo-controlled mirror used for frequency scanning. To compensate this we chose to actively stabilize the length of the cavity to an absolute frequency reference. For this we use 767\,nm laser light from a grating stabilized diode laser (the same used for trapping and cooling of ultracold potassium atoms). A couple of mW are passed through a heated potassium vapour cell and by means of modulation transfer spectroscopy (MTS)~\cite{Shirley1982, Camy1982} we obtain an error signal for locking the laser to the potassium transition. This locking technique has the advantage of suppressing open transitions and crossover resonances while enhancing the cycling transitions, thus leading to a stronger and more robust error signal~\cite{McCarron2008}. The measured locked 767\,nm laser linewidth is $120\pm 20\,$kHz. From this laser we pick off a few milliwatts which is then passed through a double-pass acousto-optical frequency shifter (Brimrose TEF-600-400) and an electro-optical modulator, driven with frequencies of approx $800\,$MHz and $6\,$MHz respectively, to generate carrier and sideband components. The radio-frequency signal applied to the frequency shifter is generated by a direct digital synthesizer (DDS) (Analog Devices AD9914) which can be tuned over a wide range. This light is then coupled into the reference cavity to generate an error signal using the Pound-Drever-Hall technique with a feedback bandwidth of approximatively $10$\,Hz. The measured tuning range of the whole system (including the frequency shifter diffraction efficiency) is $700\,$MHz which is close to the free spectral range of the reference cavity. In a related experiment we have achieved an even wider tuning range and lower power requirements by using a fiber based EOM in place of the Brimrose frequency shifter. 

\subsection{Rydberg spectroscopy of ultracold potassium}

\begin{figure}[t!]
	\centering
	\includegraphics[width = 1\columnwidth]{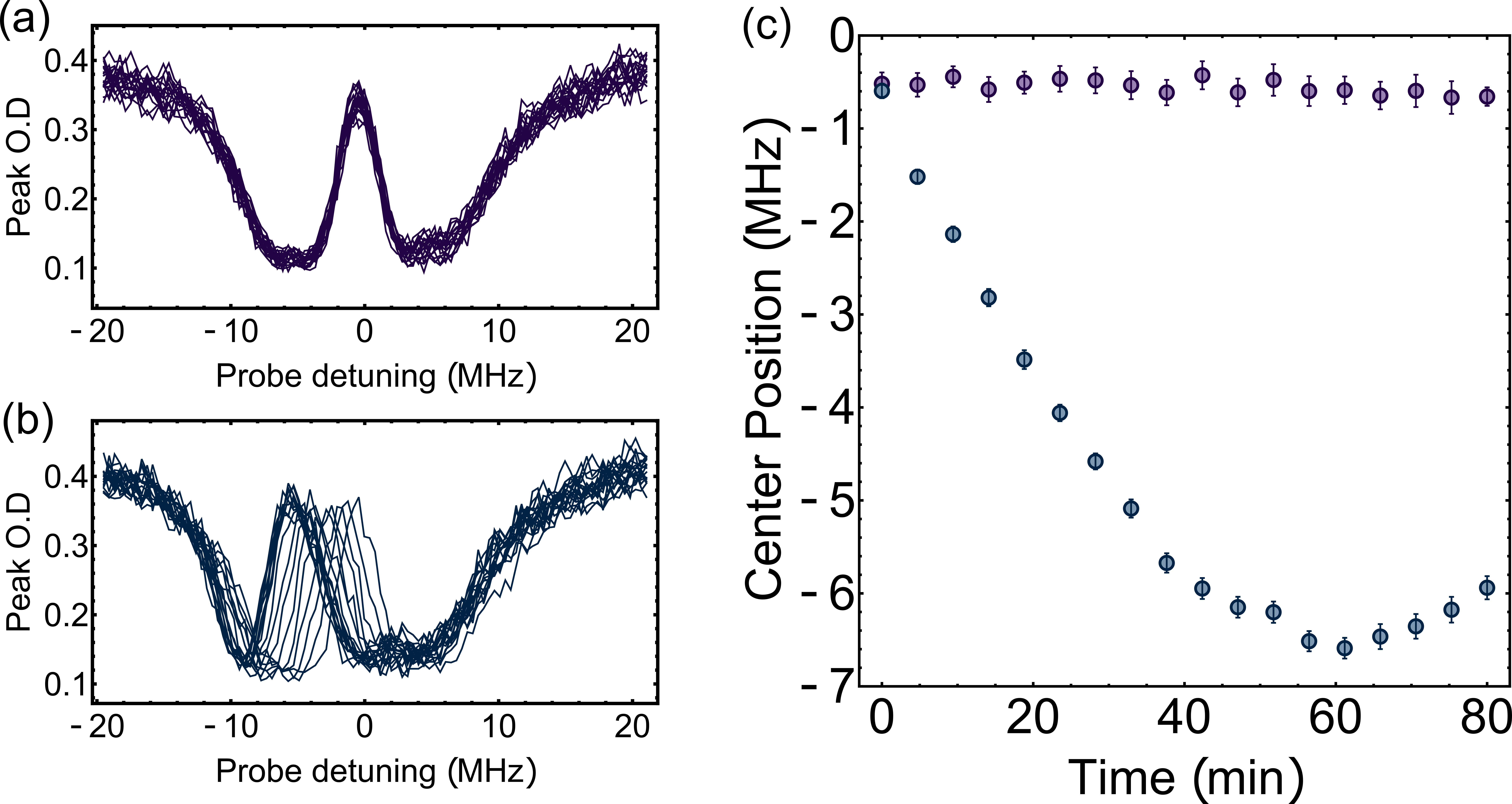}
	\caption{Frequency drift with and without cavity locking. An electromagnetically induced transparency (EIT) resonance for the $35s_{1/2}$ state of ultracold potassium provides a reference for monitoring the laser frequency stability over long time scales. (a) EIT spectra taken with the reference cavity locked to the 767\,nm laser. (b) EIT spectra without drift compensation. (c) Center position of the EIT feature recorded over 80 minutes with(purple) and without(blue) drift compensation.}
	\label{fig:LongTermStabilization}
\end{figure}

In the following we demonstrate the applicability of our laser system for atomic physics by performing two spectroscopy experiments on ultracold $^{39}$K Rydberg atoms. We start by briefly describing the apparatus used for preparing clouds of ultracold potassium atoms. Laser cooling is performed on the $4s_{1/2}\rightarrow 4p_{3/2}$ transition (D2-line) using a 767 nm laser diode. Approximately 15 mW of its total power is injected into a homebuilt tapered amplifier (TA) system which outputs more than $1$\,W of light. Subsequently this is split into three parts, which are frequency shifted using acousto-optical modulators before being injected into a second set of TA systems. The frequency shifted and amplified light is then fiber coupled to the experiment where it is used for a combined two-dimensional (2D) and three-dimensional (3D) magneto-optical trap (MOT) setup.

A typical experimental sequence is as follows. First, potassium atoms from an alkali-metal dispenser are cooled and directed from the 2D-MOT toward the 3D-MOT. After loading the 3D-MOT for a duration of 500\,ms we ramp up the magnetic quadrupole field and increase the laser detuning to compress the atom cloud. At this point the atoms are either transferred to a crossed optical dipole trap, where they are held for a short time before Rydberg excitation to observe electromagnetically induced transparency (EIT) resonances (experiment A)~\cite{Abel2009}, or we directly apply the Rydberg excitation lasers during the compressed MOT phase to induce atom losses (experiment B).

\begin{figure*}[!t]
	\centering
	\includegraphics[width = 2.0\columnwidth]{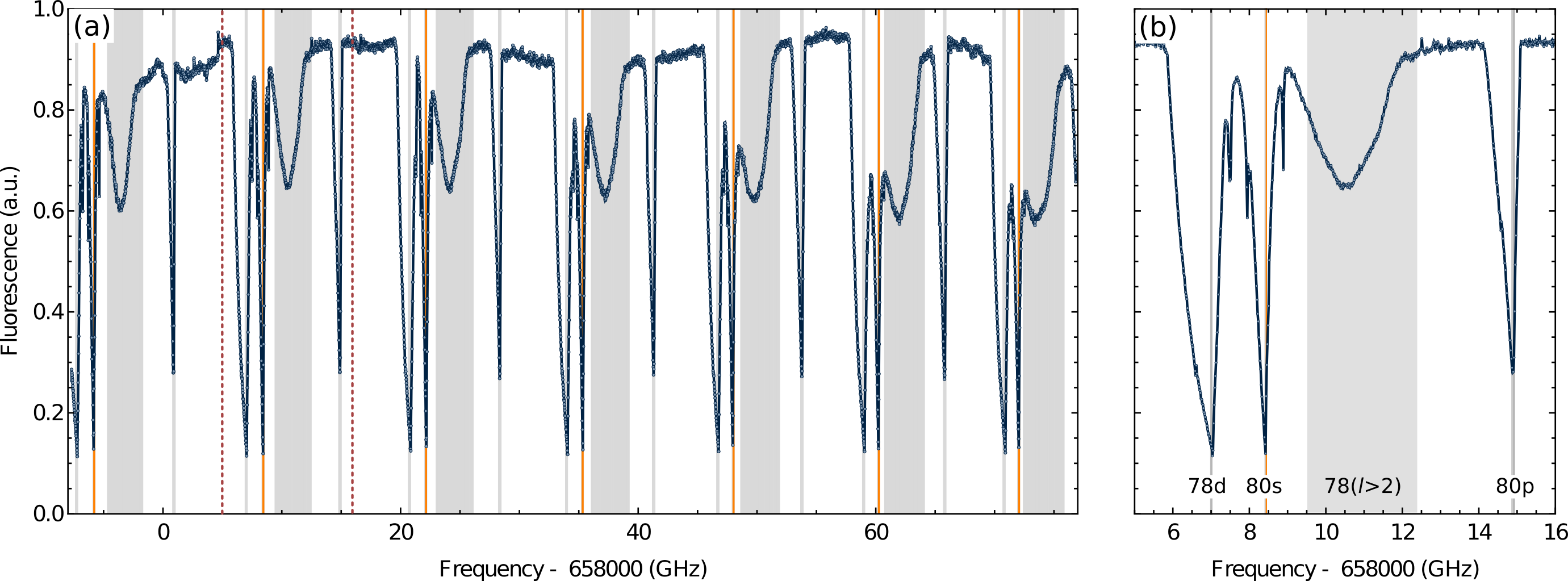}
	\caption{Rydberg loss spectrum of $^{39}$K in a magneto-optical trap. (a) The spectrum presents the allowed transitions of ns and nd states (from n=77d to n=85s) plus resonances corresponding to np-states and the hydrogenic manyfold  $l>2$ which are present due to a residual electric field. The vertical yellow and gray lines are the energy position calculated from the quantum defects in ref.~\cite{Lorenzen1983}. The spectrum is plotted around a wavelength of 455.583\,nm, for further details on the experiment see the text. (b) Zoom in of a group of s-,p-,d- and  $l>2$-states(red dashed line in (a)), showing the typical line shapes of each l-state. The narrow features on the side of the resonances are explained by the detuned light of the repumper and cooler present in the MOT.}
	\label{fig:The spectrum}
\end{figure*}

In experiment A we test the frequency stability of the laser by performing EIT measurements on atoms released from the optical dipole trap (Fig.~\ref{fig:LongTermStabilization}). The EIT measurement involves keeping the blue laser (coupling field) frequency fixed close to the $4p_{3/2}\rightarrow 35s_{1/2}$ transition while we scan the red laser (probe field) across the $4s_{1/2}\rightarrow 4p_{3/2}$ transition. On two-photon resonance we expect the appearance of a narrow transmission feature which indicates the presence of a long-lived dark state in the system. Figure~\ref{fig:LongTermStabilization}a shows a series of scans of the absorption profile taken over a time period of approximately 80 minutes. The recorded spectra have the characteristic EIT lineshape, exhibiting a clear EIT transmission peak on resonance with a subnatural width (for the probe transition) of approximately $4\,$MHz ( limited by the upper transition Rabi frequency), and with an EIT contrast larger than $80$\%. This high contrast confirms relatively good coherence of the laser excitation. The high degree of reproducibility of the spectrum over the long measurement time confirms that drifts of the experiment are well controlled. Over 80 minutes the standard deviation of the center position of the EIT feature is $\lesssim 42\,$kHz. For comparison we show a second measurement performed with the drift compensation of the reference cavity switched off (Fig.~\ref{fig:LongTermStabilization}b). We observe a systematic drift of the position of the EIT resonance as a function of time with a frequency excursion of approximately 6\,MHz. The characteristic time scale of the frequency drift is compatible with independent measurements of the temperature fluctuations in the laboratory of the order of $0.5^\circ$C.

In experiment B we test the wavelength tunability by recording Rydberg excitation spectra in a MOT for a series of principal quantum numbers. In this case the first excitation step ($4s_{1/2}\rightarrow 4p_{3/2}$ transition) is made by the MOT cooling laser which has a fixed detuning, while the frequency of the blue laser is slowly scanned via the piezo control etalon controlling the Ti:Sa frequency without the reference cavity lock (free running mode). The blue laser is focused to a waist of approximately $30\,\mu$m in the central region of the compressed MOT and has a peak intensity of $100\,$kW/cm$^2$. When the laser is tuned in resonance with a particular Rydberg level we observe a sudden drop in the MOT fluorescence. Therefore, we perform a spectroscopy measurement in which we pulse on blue laser for a duration of 20\,ms before recording the remaining MOT fluorescence level on a CCD (exposure time of 2\,ms). This sequence is repeated for many different wavelengths of the blue laser, each time with a freshly loaded MOT (Fig.~\ref{fig:The spectrum}a). The data set consists of $4900$ independent frequencies spanning approximately 80 GHz taken over a $3$ hour period. This spans the expected range from the $77d$ to the $85s$ Rydberg states.

We observe a repeated loss spectrum consisting of a series of narrow and broad loss features, some of which correspond to the expected positions of the dipole-allowed $ns$ and $nd$ states (using quantum defects taken from ref.~\cite{Lorenzen1983}). The full scan range exceeds that of the piezo control etalon alone, therefore we record the spectrum in three segments and stitch them together by post-calibrating the frequency axis of the spectrum with the known positions of the $ns$-states (vertical orange lines) and a precision wavemeter. A zoom-in of the spectrum is shown in Fig.~\ref{fig:The spectrum}b. Either side of the main resonances we observe narrow features with a minimum spectral width comparable to the size of the steps of $\approx 16\,$MHz. These sidebands can be explained by the cooler and repumper laser fields which are present in the MOT phase leading to additional resonances with the expected frequency separation of  462\,MHz. The spectrum also contains additional broad resonances which coincide closely with the expected positions of $np$-states and the hydrogenic manifold $l>2$. We conclude that these otherwise forbidden transitions are visible in our experiment due to a small residual electric field in the vacuum chamber which mixes states with different parity. This is confirmed by a calculation of the Rydberg energy spectrum assuming an electric field of 125\,mV/cm (vertical grey bands) which is consistent with the observed width of the resonances we attribute to the hydrogenic manifolds.  However we note that the $l>2$ feature is systematically shifted from the theoretical expectation, most likely due to the $l$-dependent mixing of dipole-allowed states. 

The spectral widths of $ns, np$ and $nd$ features are much broader than can be explained by the presence of the small electric field alone and they have a characteristic asymmetric shape (e.g., in fig~\ref{fig:The spectrum}b the resonances at 658\,007\,GHz and 658\,015\,GHz). We understand this as a consequence of the formation of a positively charged ultracold plasma, after the initial Rydberg excitation, associated with the continuous laser coupling on the millisecond timescale~\cite{weber2012continuous,robert2013spontaneous}. Furthermore, the asymmetry and different degrees of broadening of the resonances can be explained by the DC polarizabilities of the $ns,np$ and $nd$ states which are all negative, while the polarizability of the $nd$ states is approximately five times larger than for the $ns$ or $np$ states.

\subsection{Conclusion}
In summary, we have presented the setup and characterization of a versatile frequency doubled laser system operating at wavelengths around 460\,nm. Its key features are very high output power, wide wavelength tuning range, narrow linewidth and low intensity and frequency noise. Using this laser we have demonstrated two photon laser excitation of ultracold potassium atoms under electromagnetically induced transparency and in conditions leading to the formation of an ultracold plasma. We point out that the combination of a small electric field and the highly nonlinear loss mechanism provides a novel way to identify and measure Rydberg energy spectra for a wide range of principal quantum numbers and orbital angular momenta with a single laser system. Recently we used this laser system to reveal the influence of many-body correlations in driven-dissipative gases of Rydberg atoms~\cite{helmrich2016scaling}. It is also ideally suited for reaching conditions of strong atom-light coupling, e.g., in Rydberg dressing where long-lived admixtures of Rydberg and ground states are required. With straightforward modifications (e.g., different nonlinear crystals) this laser system can also be applied to different wavelength ranges suitable for a wide variety of atomic species. 

\acknowledgments{This work is supported in part by the Heidelberg Center for Quantum Dynamics, the European Union H2020 FET Proactive project RySQ (grant N. 640378), the DFG Collaborative Research Centre ``SFB 1225 (ISOQUANT)'', the Deutsche Forschungsgemeinschaft under WH141/1-1. AA acknowledges support by the Heidelberg Graduate School of Fundamental Physics. SH thanks the the Carl-Zeiss foundation for its financial support.}

\end{document}